\documentclass[publicdomain]{eptcs}
\usepackage{breakurl}

\title{Token-passing Optimal Reduction with Embedded Read-back}
\author{Anton Salikhmetov
\email{anton.salikhmetov@gmail.com}}
\def\titlerunning{Token-passing Optimal Reduction with Embedded Read-back}
\def\authorrunning{Anton Salikhmetov}

\usepackage[utf8]{inputenc}
\usepackage[english]{babel}
\usepackage[title]{appendix}
\usepackage{amsmath}
\usepackage{amssymb}
\usepackage{amsthm}
\usepackage{fixmath}
\usepackage{wrapfig}
\usepackage{caption}
\usepackage{subcaption}
\usepackage{tikz}
\usepackage{tikz-inet}

\DeclareFontFamily{U}{mathb}{\hyphenchar\font45}
\DeclareFontShape{U}{mathb}{m}{n}{
<-6> mathb5
<6-7> mathb6
<7-8> mathb7
<8-9> mathb8
<9-10> mathb9
<10-12> mathb10
<12-> mathb12
}{}
\DeclareSymbolFont{mathb}{U}{mathb}{m}{n}
\DeclareMathSymbol{\righttoleftarrow}{\mathrel}{mathb}{"FD}

\tikzset{every node/.style = {node distance=0em, scale=0.8}}

\newcommand{\wait}{\text{\textsf{wait}}}
\newcommand{\amb}{\text{\textsf{amb}}}
\newcommand{\hold}{\text{\textsf{hold}}}
\newcommand{\call}{\text{\textsf{call}}}
\newcommand{\eval}{\text{\textsf{eval}}}
\newcommand{\decide}{\text{\textsf{decide}}}
\newcommand{\ar}{\text{\textsf{ar}}}
\newcommand{\fv}{\text{\textsf{FV}}}

\newtheorem*{conjecture}{Conjecture}

\begin{document}
\maketitle

\begin{abstract}
We introduce a new interaction net implementation of optimal reduction for the pure untyped lambda calculus.
Unlike others, our implementation allows to reach normal form regardless of the interaction net reduction strategy using the approach of so-called token-passing nets and a non-deterministic extension for interaction nets.
Another new feature is the read-back mechanism implemented without leaving the formalism of interaction nets.
\end{abstract}

\section{Introduction}

Optimal reduction for the $\lambda$-calculus was defined by L\'evy~\cite{levy}.
One of its implementations is done by Lamping~\cite{lamping}.
Independently in~\cite{lafont}, Lafont introduced interaction nets which are graph rewriting systems similar to those used by Lamping.
Later, Lamping's optimal algorithm was redefined using the formalism of interaction nets; see \cite{optimal} where the problem of optimal reduction is covered in great detail.

Interaction nets were found as capable to encode reduction strategies directly~\cite{strategies}.
Also, a new approach of token-passing nets was introduced and demonstrated to implement call-by-value and call-by-name evaluation of $\lambda$-terms using pure interaction nets thanks to Sinot~\cite{sinot1}.
Finally, he achieved a token-passing net implementation of call-by-need evaluation~\cite{sinot2} by leaving the formalism of interaction nets.
Almeida, Pinto, and Vilaça analyzed and generalized Sinot's approach to a wider class of systems~\cite{tokenpassing}, and also stated a (currently still open) question if token-passing nets are applicable to closed reduction~\cite{closed}.

This paper is a part of an ongoing work\footnote{
The latest stable version of our implementation
is available online at \url{https://codedot.github.io/lambda/} and works in modern Web browsers
with no server side, computation being performed solely on the client side.
} on implementation of the pure untyped $\lambda K$-calculus.
For that purpose, we have introduced a domain-specific language based on the interaction calculus~\cite{calculus} with a non-deterministic extension~\cite{amb} and side effects.
However, our software implementation of this language has some limitations implied by eventually aiming at a distributed computation network.
Specifically, we have chosen not to support any specific interaction net reduction strategy as different computation nodes are meant to work independently with as minimal synchronization to each other as possible.

Hence, the optimal algorithm and many other interaction net implementations of the $\lambda$-calculus are not possible in our setting as they rely upon weak strategies aiming to reach interface-normal form of a net.
Also, they often require external garbage collection which we would rather prefer to delegate to interaction nets themselves.
Still, we found that token-passing nets fit us perfectly.
In particular, we have successfully adapted~\cite{mlc} Sinot's call-by-need for free~\cite{sinot2}.
Our current work is mostly based on the latter adaptation, thus we also base our system on interaction nets with non-deterministic extension.

In this paper, we apply the approach of token-passing nets to optimal reduction of pure untyped $\lambda K$-terms.
Our optimal implementation has two new features: reaching normal form (if any) regardless of the interaction net reduction strategy and producing textual representation of the normal form.
The first feature is in contrast with other implementations that require avoiding interactions in disconnected parts of a net.
The second feature makes more sophisticated use of \textsf{eval} and \textsf{return} tokens which Sinot's token-passing nets are based on.

\section{Preliminaries}

This section gives a rather informal brief introduction to interaction nets and their textual representation called the interaction calculus~\cite{calculus}.
Also, we will give an overview of our programming language based on the interaction calculus which we use for software experiments with interaction nets.

\begin{figure}
\centering
\caption{Primitives}
\label{primfig}
\begin{subfigure}[t]{0.3\textwidth}
\centering
\caption{Agent}
\label{primagentfig}
$$
\begin{tikzpicture}[baseline=(i.base)]
\inetcell(a){$\phantom X\alpha\phantom X$}[R]
\inetwirefree(a.pal)
\inetwirefree(a.left pax)
\inetwirefree(a.right pax)
\node (0) [right=of a.above pal] {$x_0$};
\node (1) [left=of a.above left pax] {$x_1$};
\node (i) [left=of a.above middle pax] {$\vdots$};
\node (n) [left=of a.above right pax] {$x_n$};
\end{tikzpicture}
$$
\end{subfigure}
\hfill
\begin{subfigure}[t]{0.2\textwidth}
\centering
\caption{Wiring}
\label{primwirefig}
$$
\begin{tikzpicture}[baseline=(w)]
\matrix[row sep=1.5em]{
\node (l) {$x_1$}; &
\node (i) {$\cdots$}; &
\node (r) {$x_{2k}$}; \\
\node (l') {$\phantom{x_1}$}; &
\node (w) {$\omega$}; &
\node (r') {$\phantom{x_{2k}}$}; \\
};
\inetbox{(l') (w) (r')}(b)
\inetwirecoords(l)(intersection cs:
first line={(b.north west)--(b.north east)},
second line={(l)--(l')})
\inetwirecoords(r)(intersection cs:
first line={(b.north west)--(b.north east)},
second line={(r)--(r')})
\end{tikzpicture}
$$
\end{subfigure}
\hfill
\begin{subfigure}[t]{0.4\textwidth}
\centering
\caption{Tree}
\label{primtreefig}
$$
\begin{tikzpicture}[baseline=(a)]
\inetcell[double](a){$\phantom x t\phantom x$}[U]
\inetwirefree(a.pal)
\inetwirefree(a.left pax)
\inetwirefree(a.right pax)
\node [below=of a.middle pax] {$\cdots$};
\end{tikzpicture}
\equiv
\begin{tikzpicture}[baseline=(a)]
\matrix[row sep=2.5em]{
\inetcell(a){$\phantom x \alpha \phantom x$}[U] \\
\node (i) {$\cdots$} ; \\ };
\inetcell[double, left=of i](t1){$\ t_1\ $}[U]
\inetcell[double, right=of i](tn){$\ t_n\ $}[U]
\inetwirefree(a.pal)
\inetwire(a.left pax)(t1.pal)
\inetwire(a.right pax)(tn.pal)
\inetwirefree(t1.left pax)
\inetwirefree(t1.right pax)
\node [below=of t1.middle pax] {$\cdots$};
\node (it) [below=of a.middle pax] {$\cdots$};
\inetwirefree(tn.left pax)
\inetwirefree(tn.right pax)
\node [below=of tn.middle pax] {$\cdots$};
\end{tikzpicture}
\text{ or }
\begin{tikzpicture}[baseline=(a)]
\inetcell[double,opacity=0](a){$t$}[U]
\inetwirecoords(a.above pal)(a.middle pax)
\node (i) [below=of a.middle pax] {$x$};
\end{tikzpicture}
$$
\end{subfigure}
\end{figure}
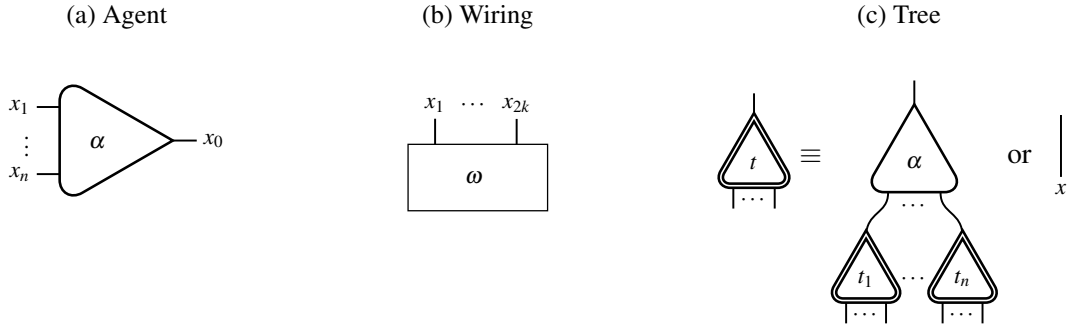

Interaction nets are graph-like structures consisting of primitives shown in Figure~\ref{primfig}.
\textit{Agents} of type $\alpha$ can be graphically represented as shown in Figure~\ref{primagentfig}.
Agents have \textit{arity} $\ar(\alpha) \ge 0$.
If $\ar(\alpha) = n$, the agent $\alpha$ has $n$ \textit{auxiliary ports} $x_1,\dots, x_n$ in addition to its \textit{principal port} $x_0$.
All agent types belong to a set $\Sigma$ called \textit{signature}.
Any port can be connected to at most one edge and ports not connected to any edge are called \textit{free ports}, the latter ones forming the \textit{interface} of an interaction net.
\textit{Wiring} $\omega$ on Figure~\ref{primwirefig} consists solely of edges.
Inductively defined \textit{trees} on  Figure~\ref{primtreefig} correspond to \textit{terms} $t ::= \alpha(t_1,\dots, t_n)\ |\ x$ in the interaction calculus, where $x$ is called a \textit{name}.

Any net $N$ can be redrawn using the previously defined wiring and tree primitives as follows:
$$
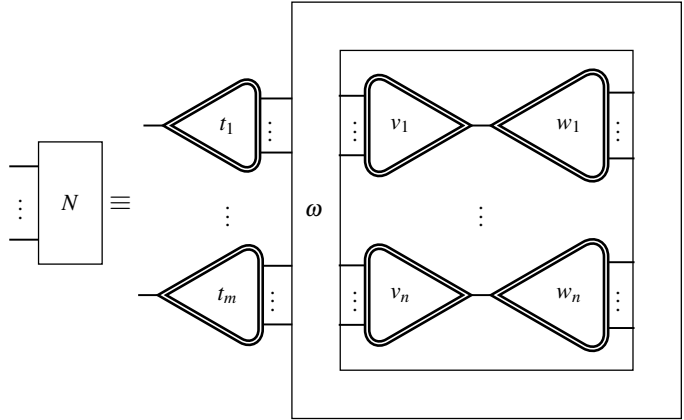
\begin{tikzpicture}[baseline=(n.base)]
\matrix{
\node (ul) {$\phantom x$}; & &
\node (u) {$\phantom x$}; \\
& \node {$\vdots\quad$}; &
\node (n) {$N$}; \\
\node (dl) {$\phantom x$}; & &
\node (d) {$\phantom x$}; \\
};
\inetbox{(d) (u)}(b)
\inetwirecoords(ul)(intersection cs:
first line={(b.north west)--(b.south west)},
second line={(ul)--(u)})
\inetwirecoords(dl)(intersection cs:
first line={(b.north west)--(b.south west)},
second line={(dl)--(d)})
\end{tikzpicture}
\equiv
\begin{tikzpicture}[baseline=(wl)]
\matrix{
& \node (wnwo) {}; & & & & & & & \\
& \node {$\phantom\omega$}; &
& \node (wnw) {$\phantom\omega$}; & & & & \\
\inetcell[double](t1){$\ \ t_1\ \ $}[L] & & & &
\inetcell[double](v1){$\ \ v_1\ \ $}[R] & &
\inetcell[double](w1){$\ \ w_1\ \ $}[L] & & \\
\node (ti) {$\vdots$}; & &
\node (wl) {$\omega$}; & & &
\node (ei) {$\vdots$}; & & &
\node (wr) {$\phantom\omega$}; \\
\inetcell[double](tm){$\ \ t_m\ \ $}[L] & & & &
\inetcell[double](vn){$\ \ v_n\ \ $}[R] & &
\inetcell[double](wn){$\ \ w_n\ \ $}[L] & \\
& & & & & & & \node (wse) {$\phantom\omega$}; & \\
& & & & & & & & \node (wseo) {$\phantom\omega$}; \\
};
\inetbox[inner sep=3em]{(wnw) (wse)}(bo)
\inetbox{(wnw) (wse)}(bi)
\inetwirefree(t1.pal)
\inetwirefree(tm.pal)
\inetwire(v1.pal)(w1.pal)
\inetwire(vn.pal)(wn.pal)
\node [right=of t1.middle pax] {$\vdots$};
\node [right=of tm.middle pax] {$\vdots$};
\node [left=of v1.middle pax] {$\vdots$};
\node [left=of vn.middle pax] {$\vdots$};
\node [right=of w1.middle pax] {$\vdots$};
\node [right=of wn.middle pax] {$\vdots$};
\inetwirecoords(t1.right pax)(intersection cs:
first line={(bo.north west)--(bo.south west)},
second line={(t1.right pax)--(t1.above right pax)})
\inetwirecoords(t1.left pax)(intersection cs:
first line={(bo.north west)--(bo.south west)},
second line={(t1.left pax)--(t1.above left pax)})
\inetwirecoords(tm.right pax)(intersection cs:
first line={(bo.north west)--(bo.south west)},
second line={(tm.right pax)--(tm.above right pax)})
\inetwirecoords(tm.left pax)(intersection cs:
first line={(bo.north west)--(bo.south west)},
second line={(tm.left pax)--(tm.above left pax)})
\inetwirecoords(w1.right pax)(intersection cs:
first line={(bi.north east)--(bi.south east)},
second line={(w1.right pax)--(w1.above right pax)})
\inetwirecoords(w1.left pax)(intersection cs:
first line={(bi.north east)--(bi.south east)},
second line={(w1.left pax)--(w1.above left pax)})
\inetwirecoords(wn.right pax)(intersection cs:
first line={(bi.north east)--(bi.south east)},
second line={(wn.right pax)--(wn.above right pax)})
\inetwirecoords(wn.left pax)(intersection cs:
first line={(bi.north east)--(bi.south east)},
second line={(wn.left pax)--(wn.above left pax)})
\inetwirecoords(v1.right pax)(intersection cs:
first line={(bi.north west)--(bi.south west)},
second line={(v1.right pax)--(v1.above right pax)})
\inetwirecoords(v1.left pax)(intersection cs:
first line={(bi.north west)--(bi.south west)},
second line={(v1.left pax)--(v1.above left pax)})
\inetwirecoords(vn.right pax)(intersection cs:
first line={(bi.north west)--(bi.south west)},
second line={(vn.right pax)--(vn.above right pax)})
\inetwirecoords(vn.left pax)(intersection cs:
first line={(bi.north west)--(bi.south west)},
second line={(vn.left pax)--(vn.above left pax)})
\end{tikzpicture}
$$
which in the interaction calculus corresponds to a \textit{configuration} $\langle t_1, \dots, t_m\ |\ v_1 = w_1, \dots,\ v_n = w_n \rangle$, where $t_i$, $v_i$, and $w_i$ are arbitrary terms.
The ordered sequence $t_1,\dots, t_m$ in the left-hand side is called an \textit{interface}, while the right-hand side contains an unordered multiset of \textit{equations} $v_i = w_i$.
The wiring $\omega$ translates to names, and each name has to occur exactly twice in a configuration.

Just like in the $\lambda$-calculus, the interaction calculus has the notions of $\alpha$-conversion and substitution naturally defined on configurations.
Specifically, both occurrences of any name can be replaced with a new name if the latter does not occur in a given configuration.
In turn, \textit{substitution} $t[x := u]$ is the result of replacing the name $x$ in term $t$ with another term $u$ if $x$ has exactly one occurrence in term $t$.

When two agents are connected to each other with their principal ports, they form an \textit{active pair}.
For active pairs one can introduce \textit{interaction rules} which describe how the active pair rewrites to another net.
Graphically, any interaction rule can be represented as follows:
$$
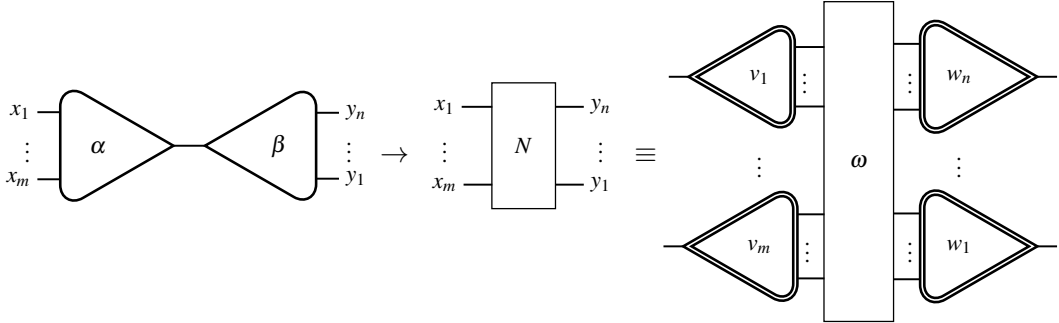
\begin{tikzpicture}[baseline=(yi.base)]
\matrix[column sep=1em]{
\inetcell(a){$\phantom X\alpha\phantom X$}[R] &
\inetcell(b){$\phantom X\beta\phantom X$}[L] \\ };
\inetwirefree(a.left pax)
\inetwirefree(a.right pax)
\inetwire(a.pal)(b.pal)
\inetwirefree(b.left pax)
\inetwirefree(b.right pax)
\node (x1) [left=of a.above left pax] {$x_1$};
\node (xi) [left=of a.above middle pax] {$\vdots$};
\node (xn) [left=of a.above right pax] {$x_m$};
\node (y1) [right=of b.above left pax] {$y_1$};
\node (yi) [right=of b.above middle pax] {$\vdots$};
\node (yn) [right=of b.above right pax] {$y_n$};
\end{tikzpicture}
\rightarrow
\begin{tikzpicture}[baseline=(xi.base)]
\matrix[column sep=1.5em]{
\node (x1) {$x_1$}; &
\node (t) {$\phantom x$}; &
\node (yn) {$y_n$}; \\
\node (xi) {$\vdots$}; &
\node (n) {$N$}; &
\node (yi) {$\vdots$}; \\
\node (xn) {$x_m$}; &
\node (b) {$\phantom x$}; &
\node (y1) {$y_1$}; \\ };
\inetbox{(b) (t)}(box)
\inetwirecoords(x1)(intersection cs:
first line={(box.north west)--(box.south west)},
second line={(x1)--(yn)})
\inetwirecoords(xn)(intersection cs:
first line={(box.north west)--(box.south west)},
second line={(xn)--(y1)})
\inetwirecoords(yn)(intersection cs:
first line={(box.north east)--(box.south east)},
second line={(x1)--(yn)})
\inetwirecoords(y1)(intersection cs:
first line={(box.north east)--(box.south east)},
second line={(xn)--(y1)})
\end{tikzpicture}
\equiv
\begin{tikzpicture}[baseline=(wl)]
\matrix[column sep=1.5em]{
& \node (wt) {$\phantom\omega$}; & \\
\inetcell[double](t1){$\ \ v_1\ \ $}[L] & &
\inetcell[double](v1){$\ \ w_n\ \ $}[R] \\
\node (ti) {$\vdots$}; &
\node (wl) {$\omega$}; &
\node (ei) {$\vdots$}; \\
\inetcell[double](tm){$\ \ v_m\ \ $}[L] & &
\inetcell[double](vn){$\ \ w_1\ \ $}[R] \\
& \node (wb) {$\phantom\omega$}; & \\
};
\inetbox{(wt) (wb)}(b)
\inetwirefree(t1.pal)
\inetwirefree(tm.pal)
\inetwirefree(v1.pal)
\inetwirefree(vn.pal)
\node [right=of t1.middle pax] {$\vdots$};
\node [right=of tm.middle pax] {$\vdots$};
\node [left=of v1.middle pax] {$\vdots$};
\node [left=of vn.middle pax] {$\vdots$};
\inetwirecoords(t1.right pax)(intersection cs:
first line={(b.north west)--(b.south west)},
second line={(t1.right pax)--(t1.above right pax)})
\inetwirecoords(t1.left pax)(intersection cs:
first line={(b.north west)--(b.south west)},
second line={(t1.left pax)--(t1.above left pax)})
\inetwirecoords(tm.right pax)(intersection cs:
first line={(b.north west)--(b.south west)},
second line={(tm.right pax)--(tm.above right pax)})
\inetwirecoords(tm.left pax)(intersection cs:
first line={(b.north west)--(b.south west)},
second line={(tm.left pax)--(tm.above left pax)})
\inetwirecoords(v1.right pax)(intersection cs:
first line={(b.north east)--(b.south east)},
second line={(v1.right pax)--(v1.above right pax)})
\inetwirecoords(v1.left pax)(intersection cs:
first line={(b.north east)--(b.south east)},
second line={(v1.left pax)--(v1.above left pax)})
\inetwirecoords(vn.right pax)(intersection cs:
first line={(b.north east)--(b.south east)},
second line={(vn.right pax)--(vn.above right pax)})
\inetwirecoords(vn.left pax)(intersection cs:
first line={(b.north east)--(b.south east)},
second line={(vn.left pax)--(vn.above left pax)})
\end{tikzpicture}
$$
where $\alpha, \beta \in \Sigma$, and the net $N$ is redrawn using primitives of wirings and trees in order to translate the rule into the interaction calculus as $\alpha[v_1, \dots, v_m] \bowtie \beta[w_1, \dots, w_n]$ using Lafont's notation.
A net with no active pairs is said to be in \textit{normal form}.
A signature $\Sigma$ (with mapping $\ar$ defined on it) along with a set of interaction rules defined for agents $\alpha \in \Sigma$ together constitute an \textit{interaction system}.

Now, let us consider an example for the notions introduced above in this section.
Figure~\ref{egfig} shows two interaction rules for commonly used agents $\epsilon$ and $\delta$ and a simple interaction net to which these interaction rules are applied.
Using Lafont's notation, the erasing rule from Figure~\ref{egerasefig} is written as $\epsilon \bowtie \alpha[\epsilon,\dots, \epsilon]$, while the duplication rule given in Figure~\ref{egdupfig} can be represented as follows:
$$
\delta[\alpha(x_1,\dots, x_n), \alpha(y_1,\dots, y_n)] \bowtie \alpha[\delta(x_1, y_1),\dots, \delta(x_n, y_n)].
$$
Figure~\ref{egloopfig} provides an example of a non-terminating net which reduces to itself.
In terms of the interaction calculus, one can write this net as a configuration $\langle \varnothing\ |\ \delta(\epsilon, x) = \gamma(x, \epsilon)\rangle$ with no interface.

\begin{figure}[b]
\centering
\caption{Example}
\label{egfig}
\begin{subfigure}[t]{0.3\textwidth}
\centering
\caption{Erasing}
\label{egerasefig}
$$
\begin{tikzpicture}[baseline=(a)]
\matrix[row sep=1em]{
\inetcell(e){$\epsilon$}[D] \\
\inetcell(a){$\phantom x\alpha\phantom x$}[U] \\ };
\node (i) [below=of a.middle pax] {$\dots$};
\inetwirefree(a.left pax)
\inetwirefree(a.right pax)
\inetwire(e.pal)(a.pal)
\end{tikzpicture}
\rightarrow
\begin{tikzpicture}[baseline=(i.base)]
\matrix{
\inetcell(1){$\epsilon$} &
\node (i) {$\dots$}; &
\inetcell(n){$\epsilon$} \\ };
\inetwirefree(1.pal)
\inetwirefree(n.pal)
\end{tikzpicture}
$$
\end{subfigure}
\hfill
\begin{subfigure}[t]{0.4\textwidth}
\centering
\caption{Duplication}
\label{egdupfig}
$$
\begin{tikzpicture}[baseline=(a)]
\matrix[row sep=1em]{
\inetcell(d){$\delta$}[D] \\
\inetcell(a){$\phantom x\alpha\phantom x$}[U] \\ };
\node (i) [below=of a.middle pax] {$\dots$};
\inetwirefree(a.left pax)
\inetwirefree(a.right pax)
\inetwirefree(d.left pax)
\inetwirefree(d.right pax)
\inetwire(d.pal)(a.pal)
\end{tikzpicture}
\rightarrow
\begin{tikzpicture}[baseline=(i.base)]
\matrix[row sep=2em]{
\inetcell(l){$\phantom x\alpha\phantom x$}[U] & &
\inetcell(r){$\phantom x\alpha\phantom x$}[U] \\
\inetcell(1){$\delta$} &
\node (i) {$\dots$}; &
\inetcell(n){$\delta$} \\ };
\node (li) [below=of l.middle pax] {$\dots$};
\node (ri) [below=of r.middle pax] {$\dots$};
\inetwire(l.left pax)(1.right pax)
\inetwire(l.right pax)(n.right pax)
\inetwire(r.left pax)(1.left pax)
\inetwire(r.right pax)(n.left pax)
\inetwirefree(1.pal)
\inetwirefree(n.pal)
\inetwirefree(l.pal)
\inetwirefree(r.pal)
\end{tikzpicture}
$$
\end{subfigure}
\hfill
\begin{subfigure}[t]{0.2\textwidth}
\centering
\caption{Non-termination}
\label{egloopfig}
$$
\begin{tikzpicture}[baseline=(g.above pal)]
\matrix[row sep=0.7em]{
\inetcell(g){$\gamma$}[D] &
\inetcell[opacity=0](gr){$\gamma$}[D] \\
\inetcell(d){$\delta$}[U] &
\inetcell[opacity=0](dr){$\delta$}[U] \\
};
\inetcell[above=of g.above right pax](et){$\epsilon$}[D]
\inetcell[below=of d.above left pax](eb){$\epsilon$}[U]
\inetwire(g.pal)(d.pal)
\inetwire(et.pal)(g.right pax)
\inetwire(g.left pax)(gr.middle pax)
\inetwire(d.right pax)(dr.middle pax)
\inetwire(eb.pal)(d.left pax)
\inetwirecoords(gr.middle pax)(dr.middle pax)
\end{tikzpicture}
\righttoleftarrow^*
$$
\end{subfigure}
\end{figure}

The interaction calculus defines reduction on configurations in more details than seen from graph rewriting defined on interaction nets.
Namely, if $\alpha[v_1, \dots, v_m] \bowtie \beta[w_1, \dots, w_n]$, the following reduction:
$$
\langle\vec t\ |\ \alpha(t_1,\dots, t_m) = \beta(u_1,\dots, u_n),\ \Delta\rangle
\rightarrow
\langle\vec t\ |\ t_1 = v_1,\dots,\ t_m = v_m,\ u_1 = w_1,\dots,\ u_n = w_n,\ \Delta\rangle
$$
is called \textit{interaction}.
When one of equations has the form of $x = u$, \textit{indirection} can be applied resulting in substitution of the other occurrence of the name $x$ in some term $t$:
$$
\langle\dots t \dots\ |\ x = u,\ \Delta\rangle
\rightarrow
\langle\dots t[x := u] \dots\ |\ \Delta\rangle
\quad \text{or} \quad
\langle\vec t\ |\ x = u,\ t = w,\ \Delta\rangle
\rightarrow
\langle\vec t\ |\ t[x := u] = w,\ \Delta\rangle.
$$
An equation $t = x$ is called a \textit{deadlock} if the name $x$ has occurrence in the term $t$.
We only consider deadlock-free nets.
Together, interaction and indirection define the reduction relation on configurations.
The fact that configuration $c$ reduces to its \textit{normal form} $c'$ with no equations left is denoted as $c \downarrow c'$.

Coming back to the example of a non-terminating net shown in Figure~\ref{egloopfig}, the infinite reduction sequence starting from the corresponding configuration in the interaction calculus is as follows:
\begin{align*}
&\langle \varnothing\ |\ \delta(\epsilon, x) = \gamma(x, \epsilon)\rangle \rightarrow \\
&\langle \varnothing\ |\ \epsilon = \gamma(x_1, x_2),\ x = \gamma(y_1, y_2),\ x = \delta(x_1, y_1),\ \epsilon = \delta(x_2, y_2)\rangle \rightarrow^* \\
&\langle \varnothing\ |\ x_1 = \epsilon,\ x_2 = \epsilon,\ x = \gamma(y_1, y_2),\ x = \delta(x_1, y_1),\ x_2 = \epsilon,\ y_2 = \epsilon\rangle \rightarrow^* \\
&\langle \varnothing\ |\ \delta(\epsilon, x) = \gamma(x, \epsilon)\rangle \rightarrow \dots
\end{align*}
In our DSL\footnote{
Our programming language is provided with a Node.js package called JavaScript Engine for Interaction Nets which is available at \url{https://www.npmjs.com/package/inet-lib} in the NPM repository.
} which is similar to the UNIX utilities \texttt{yacc(1)/lex(1)} by structure and lexically close to LaTeX representation for the interaction calculus, the example net can be specified as follows:
\begin{verbatim}
\epsilon {
        console.log("epsilon >< delta");
} \delta[\epsilon, \epsilon];

\epsilon {
        console.log("epsilon >< gamma");
} \gamma[\epsilon, \epsilon];

\delta[\gamma(x, y), \gamma(v, w)] {
        console.log("delta >< gamma");
} \gamma[\delta(x, v), \delta(y, w)];

$$

\delta(\epsilon, x) = \gamma(x, \epsilon);
\end{verbatim}

Note that our programming language allows side effects written in imperative style, thus enabling execution of arbitrary code, including input/output as well as conditional multiple rules for a given pair $\alpha_i \bowtie \beta_j$ depending on $i$ and $j$, while manipulating $i$ and $j$ as arbitrary data attached to agents.

Our implementation of this language does not rely on any external garbage collection as it does not distinguish disconnected subnets.
The latter is due to lack of interface as we specify a net by a multiset of equations only.
In order to still represent a net with a non-empty interface, one needs to modify the net, for instance, by attaching agents with zero arity to all of its free ports.

An interaction system's signature and arities of agents are derived automatically at compile time based on input interaction rules and initial configuration.
The interaction rules along with their side effects are compiled in advance before starting reduction of a net, the compiler preparing a table for a fast $O(1)$ rule search at run time.
After that, the initial configuration is pushed into a FIFO queue which is then processed by evaluator until no more equations left.

Finally, implementation of our DSL includes non-deterministic extension for interaction nets in the same form as described in Section~\ref{ambsec}, namely using McCarthy's $\amb$ agent which allows to simulate agents with multiple principal ports such as the sharing agent $s$ used by Sinot in~\cite{sinot2}.

\clearpage
\section{Non-deterministic extension}
\label{ambsec}

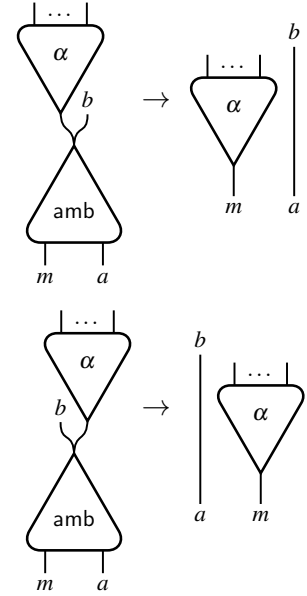
\begin{wrapfigure}{R}{0.4\textwidth}
\caption{Non-deterministic agent}
\label{ambfig}
$$
\begin{tikzpicture}[baseline=(b.base)]
\matrix[row sep=1em]{
\inetcell[opacity=0](p){$X$}[U] \\
\inetcell(a){$\amb$}[U] \\ };
\node (m) [below=of a.above right pax] {$a$};
\node (t) [below=of a.above left pax] {$m$};
\node (b) [above=of p.right pax] {$b$};
\inetcell[above=of p.left pax](x){$\phantom x\alpha\phantom x$}[D]
\node (d) [above=of x.middle pax] {$\dots$};
\inetwirefree(a.left pax)
\inetwirefree(a.right pax)
\inetwirefree(x.left pax)
\inetwirefree(x.right pax)
\inetwire(a.pal)(p.right pax)
\inetwire(a.pal)(x.pal)
\end{tikzpicture}
\rightarrow
\begin{tikzpicture}[baseline=(x)]
\matrix[row sep=1em]{
& \node (b) {$b$}; \\
\inetcell(x){$\phantom x\alpha\phantom x$}[D] & \\
\node (m) {$m$}; & \node (t) {$a$}; \\ };
\node (d) [above=of x.middle pax] {$\dots$};
\inetwirefree(x.left pax)
\inetwirefree(x.right pax)
\inetwirecoords(x.pal)(m)
\inetwirecoords(t)(b)
\end{tikzpicture}
$$
$$
\begin{tikzpicture}[baseline=(b.base)]
\matrix[row sep=1em]{
\inetcell[opacity=0](p){$X$}[U] \\
\inetcell(a){$\amb$}[U] \\ };
\node (m) [below=of a.above right pax] {$a$};
\node (t) [below=of a.above left pax] {$m$};
\node (b) [above=of p.left pax] {$b$};
\inetcell[above=of p.right pax](x){$\phantom x\alpha\phantom x$}[D]
\node (d) [above=of x.middle pax] {$\dots$};
\inetwirefree(a.left pax)
\inetwirefree(a.right pax)
\inetwirefree(x.left pax)
\inetwirefree(x.right pax)
\inetwire(a.pal)(p.left pax)
\inetwire(a.pal)(x.pal)
\end{tikzpicture}
\rightarrow
\begin{tikzpicture}[baseline=(x)]
\matrix[row sep=1em]{
\node (b) {$b$}; & \\
& \inetcell(x){$\phantom x\alpha\phantom x$}[D] \\
\node (t) {$a$}; & \node (m) {$m$}; \\ };
\node (d) [above=of x.middle pax] {$\dots$};
\inetwirefree(x.left pax)
\inetwirefree(x.right pax)
\inetwirecoords(x.pal)(m)
\inetwirecoords(t)(b)
\end{tikzpicture}
$$
\end{wrapfigure}

We work in the interaction calculus~\cite{calculus} extended with a non-deterministic agent $\amb$~\cite{amb}.
This agent will be required in the waiting construct which is described in Section~\ref{waitsec}.
Specifically, it is used to handle non-deterministic nature of deciding whether a given sub-term of an encoded $\lambda K$-term is to be reduced and erased. 
The context where we need the $\amb$ agent is similar to where multiple principal ports were necessary for token-passing net implementation of call-by-need evaluation by Sinot.

We represent this agent in a more conservative fashion than it was suggested in the original paper~\cite{amb}. Specifically, we prepend the list of auxiliary ports of $\amb$ with its extra principal port and introduce the following conversion:
$$
\langle\vec t\ |\ t = \amb(u, v, w),\ \Delta\rangle 
=
\langle\vec t\ |\ u = \amb(t, v, w),\ \Delta\rangle.
$$

We assume that any interaction system's signature $\Sigma$ is implicitly extended by $\amb$ with ${\ar(\amb) = 3}$, while its set of rules is implicitly extended with
$$
\forall \alpha \in \Sigma:
\alpha[\vec x] \bowtie \amb[y, \alpha(\vec x), y].
$$

Figure~\ref{ambfig} gives a graphical representation of $\amb$.

\section{Optimal reduction}

The pure interaction net implementation of optimal reduction upon which we base our interaction system is the main one used through most of the book by Asperti and Guerrini~\cite[pp.~39--40]{optimal}.
In that system, all agents have box level numbers attached to them. 
From the viewpoint of the interaction calculus, this fact results in an interaction system with an infinite signature and an infinite set of interaction rules.
Specifically, its signature can be defined as follows:
$$
\Sigma_O = \{\epsilon\} \cup \{\lambda_i, @_i, \delta_i, \doublecap_i, \sqcup_i\ |\ i \in \mathbb{N}\},
$$
with $\ar(\epsilon) = 0$, $\ar(\doublecap_i) = \ar(\sqcup_i) = 1$, and $\ar(\lambda_i) = \ar(@_i) = \ar(\delta_i) = 2$.
In Section~\ref{readbacksec} where we define the embedded read-back mechanism, we will extend the signature with agents that have $\lambda$-terms and contexts attached rather than natural numbers.

The set of interaction rules consists of two groups.
The first group includes \textit{annihilation} rules:
\begin{align*}
\doublecap_i[x] &\bowtie \doublecap_i[x]; \\
\sqcup_i[x] &\bowtie \sqcup_i[x]; \\
\lambda_i[x, y] &\bowtie @_i[x, y]; \\
\delta_i[x, y] &\bowtie \delta_i[x, y].
\end{align*}
The second group contains the following \textit{propagation} rules, where $\alpha_j \in \Sigma_O$ and $i < j$:
\begin{align*}
\doublecap_i[\alpha_{j - 1}(x_1, \dots, x_n)] &\bowtie \alpha_j[\doublecap_i(x_1), \dots, \doublecap_i(x_n)]; \\
\sqcup_i[\alpha_{j + 1}(x_1, \dots, x_n)] &\bowtie \alpha_j[\sqcup_i(x_1), \dots, \sqcup_i(x_n)]; \\
\delta_i[\alpha_j(x_1, \dots, x_n), \alpha_j(y_1, \dots, y_n)] &\bowtie \alpha_j[\delta_i(x_1, y_1), \dots, \delta_i(x_n, y_n)].
\end{align*}

The initial encoding of $\lambda$-terms we base our work on is the one given in~\cite[pp.~41--42]{optimal}, although other compatible versions can also be used as described in~\cite[Chapter 8]{optimal}.
We preserve both original interaction rules and original initial encoding with exceptions for the $\beta$-reduction $\lambda_i \bowtie @_i$ rule modified and free variables allowed in a given $\lambda$-term.
The former modification is to be covered in Section~\ref{waitsec} about the waiting construct, and the latter feature will be a part of the read-back mechanism embedded into our interaction system described in Section~\ref{readbacksec}.
Together, they will constitute our token-passing net implementation of optimal reduction.

In order to make it easier to work with the interaction calculus below, we will denote the (graphical) encoding $[M]$ of a given $\lambda$-term $M$ as (textual) configuration ${\langle x\ |\ [M, x]\rangle}$, where $x$ is a name for the only free port in its interface, and $[M, x]$ is a multiset of equations that correspond to the initial encoding $[M]$.

\section{Waiting construct}
\label{waitsec}

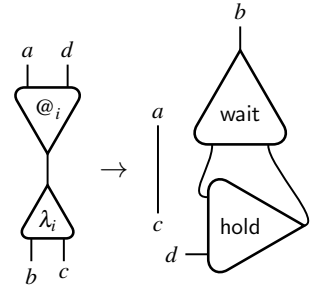
\begin{wrapfigure}{R}{0.36\textwidth}
\caption{Waiting construct}
\label{waitfig}
$$
\begin{tikzpicture}[baseline=(appl.above pal)]
\matrix[row sep=1em,ampersand replacement=\&]{
\inetcell(appl){$@_i$}[D] \\
\inetcell(abst){$\lambda_i$}[U] \\ };
\node (a) [above=of appl.above right pax] {$a$};
\node (b) [below=of abst.above left pax] {$b$};
\node (c) [below=of abst.above right pax] {$c$};
\node (d) [above=of appl.above left pax] {$d$};
\inetwirefree(appl.left pax)
\inetwirefree(appl.right pax)
\inetwirefree(abst.left pax)
\inetwirefree(abst.right pax)
\inetwire(appl.pal)(abst.pal)
\end{tikzpicture}
\rightarrow
\begin{tikzpicture}[baseline=(appl.above pal)]
\matrix[row sep=1em,column sep=0.64em,ampersand replacement=\&]{
\node (a) {$a$}; \&
\inetcell(wait){$\wait$}[U] \\
\node (c) {$c$}; \&
\inetcell(hold){$\hold$}[R] \\ };
\node (b) [above=of wait.above pal] {$b$};
\node (d) [left=of hold.above right pax] {$d$};
\inetwirecoords(a)(c)
\inetwirefree(wait.pal)
\inetwirefree(hold.right pax)
\inetwire(wait.left pax)(hold.left pax)
\inetwire(wait.right pax)(hold.pal)
\end{tikzpicture}
$$
\end{wrapfigure}

Our main modification to the optimal algorithm on the way to its token-passing net version is to replace the $\beta$-reduction rule ${@_i[x, y] \bowtie \lambda_i[x, y]}$ with
$$
@_i[x, y] \bowtie \lambda_i[\wait(z, \hold(z, x)), y],
$$
extending the original signature $\Sigma_O$ with a set of agents used to represent what we call the \textit{waiting construct}:
$$
\Sigma_W = \Sigma_O \cup \{\wait, \hold, \decide, \eval, \call\},
$$
with ${\ar(\wait) = \ar(\hold) = \ar(\decide) = 2}$, $\ar(\eval) = 1$, and $\ar(\call) = 0$.
Figure~\ref{waitfig} provides graphical representation of the modified $\beta$-reduction rule.
Note the waiting construct appearing between variable $b$ and argument $d$ is similar to the $!$ agent used by Mackie in~\cite{strategies} to cut edges corresponding to active pairs.

\begin{figure}[h]
\centering
\begin{minipage}{0.45\linewidth}
\caption{Propagation through fan-in}
\label{dupfig}
$$
\begin{tikzpicture}[baseline=(agent)]
\matrix[row sep=1em]{
\inetcell(agent){$\delta_i$}[D] \\
\inetcell(start){$\wait$}[U] \\ };
\inetwirefree(agent.left pax)
\inetwirefree(agent.right pax)
\inetwirefree(start.left pax)
\inetwirefree(start.right pax)
\inetwire(agent.pal)(start.pal)
\end{tikzpicture}
\rightarrow
\begin{tikzpicture}[baseline=(copy)]
\matrix[row sep=1em]{
\inetcell(start1){$\wait$}[U] &
\inetcell(start2){$\wait$}[U] \\
\inetcell(copy){$\delta_i$}[D] &
\inetcell(amb){$\amb$}[U] \\
& \inetcell(wait){$\decide$}[L] \\ };
\inetwirefree(start1.pal)
\inetwirefree(start2.pal)
\inetwire(start1.left pax)(copy.right pax)
\inetwire(start2.left pax)(copy.left pax)
\inetwire(amb.pal)(start1.right pax)
\inetwire(amb.pal)(start2.right pax)
\inetwirefree(copy.pal)
\inetwirefree(wait.left pax)
\inetwire(wait.pal)(amb.left pax)
\inetwire(wait.right pax)(amb.right pax)
\end{tikzpicture}
$$
\end{minipage}
\quad
\begin{minipage}{0.45\linewidth}
\caption{Propagation through application}
\label{appfig}
$$
\begin{tikzpicture}[baseline=(agent)]
\matrix[row sep=1em]{
\inetcell(agent){$@_i$}[D] \\
\inetcell(start){$\wait$}[U] \\ };
\inetwirefree(agent.left pax)
\inetwirefree(agent.right pax)
\inetwirefree(start.left pax)
\inetwirefree(start.right pax)
\inetwire(agent.pal)(start.pal)
\end{tikzpicture}
\rightarrow
\begin{tikzpicture}[baseline=(app)]
\matrix[row sep=1em]{
\inetcell(wait){$\wait$}[U] & \\
\inetcell(app){$@_i$}[D] &
\inetcell(hold){$\hold$}[U] \\
& \inetcell(old){$\wait$}[U] \\ };
\inetwirefree(wait.pal)
\inetwire(wait.left pax)(app.right pax)
\inetwirefree(app.left pax)
\inetwire(app.pal)(hold.left pax)
\inetwire(wait.right pax)(hold.pal)
\inetwire(old.pal)(hold.right pax)
\inetwirefree(old.left pax)
\inetwirefree(old.right pax)
\end{tikzpicture}
$$
\end{minipage}
\end{figure}

The waiting construct propagates through the body of every abstraction after substitution, blocking possibly unnecessary $\beta$-redexes until they are called.
This mechanism consists of several additional interaction rules we define through the rest of this section.
Since the process of deciding whether a given redex is needed has non-deterministic nature, it is the waiting construct that requires non-deterministic extension for interaction nets we discussed earlier.
In particular, interaction between a fan-in agent (denoted below as $\delta_i$) and $\wait$ results in creation of an ambiguous $\decide$ agent with two principal ports, the latter one being simulated using $\amb$ as shown in Figure~\ref{dupfig}.

Figure~\ref{appfig} illustrates the $@_i \bowtie \wait$ rule.
Note that rather than just passing through application, the waiting construct initiates another waiting construct on the way to the root of application.
Otherwise, an $\epsilon$ agent that performs garbage collection from the root of application might be unable to reach either of application's sides.
That, in turn, could lead to a disconnected net, which would become blocked non-interacting garbage.
We avoid that thanks to the structure of our $@_i \bowtie \wait$ rule.

Unblocking evaluation happens through $\wait \bowtie \eval$ interaction which is implemented in a fashion similar to how evaluation strategies are encoded in Mackie's paper~\cite{strategies}.
Although, the corresponding interaction rules are different due to the simulation of multiple principal ports:
$$
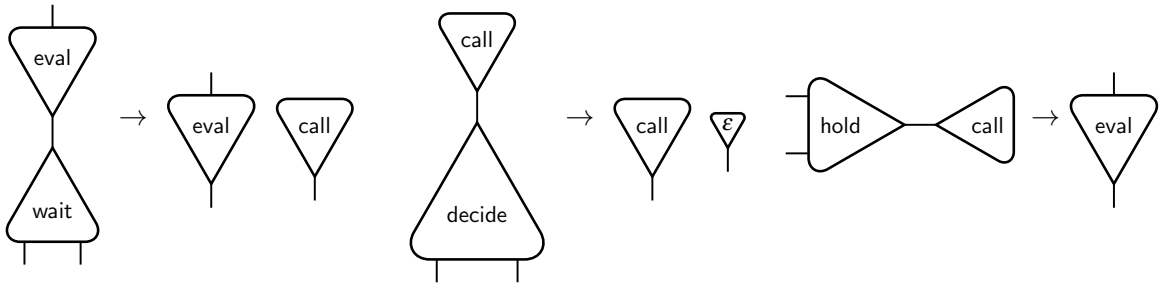
\begin{tikzpicture}[baseline=(start.above pal)]
\matrix[row sep=1em]{
\inetcell(agent){$\eval$}[D] \\
\inetcell(start){$\wait$}[U] \\ };
\inetwirefree(agent.middle pax)
\inetwirefree(start.left pax)
\inetwirefree(start.right pax)
\inetwire(agent.pal)(start.pal)
\end{tikzpicture}
\rightarrow
\begin{tikzpicture}[baseline=(agent)]
\matrix[row sep=1em, column sep=0.5em]{
\inetcell(agent){$\eval$}[D] &
\inetcell(go){$\call$}[D] \\ };
\inetwirefree(agent.middle pax)
\inetwirefree(agent.pal)
\inetwirefree(go.pal)
\end{tikzpicture}
\quad
\begin{tikzpicture}[baseline=(wait.pal)]
\matrix[row sep=1em, column sep=0.5em]{
\inetcell(go){$\call$}[D] \\
\inetcell(wait){$\decide$}[U] \\ };
\inetwirefree(wait.left pax)
\inetwirefree(wait.right pax)
\inetwire(wait.pal)(go.pal)
\end{tikzpicture}
\rightarrow
\begin{tikzpicture}[baseline=(go)]
\matrix[row sep=1em, column sep=0.5em]{
\inetcell(go){$\call$}[D] &
\inetcell(wait){$\epsilon$}[D] \\ };
\inetwirefree(go.pal)
\inetwirefree(wait.pal)
\end{tikzpicture}
\quad
\begin{tikzpicture}[baseline=(wait)]
\matrix[row sep=2em, column sep=1em]{
\inetcell(wait){$\hold$}[R] &
\inetcell(go){$\call$}[L] \\ };
\inetwirefree(wait.left pax)
\inetwirefree(wait.right pax)
\inetwire(wait.pal)(go.pal)
\end{tikzpicture}
\rightarrow
\begin{tikzpicture}[baseline=(eval)]
\inetcell(eval){$\eval$}[D]
\inetwirefree(eval.pal)
\inetwirefree(eval.middle pax)
\end{tikzpicture}
$$

The rest of propagation and garbage collection interaction rules are more or less straightforward, so we use Lafont's notation instead to put them all together:
\begin{align*}
\eval[\lambda_i(x, y)] &\bowtie \lambda_i[x, \eval(y)]; \\
\eval[\delta_i(x, y)] &\bowtie \delta_i[x, y]; \\
\eval[x] &\bowtie \wait[\eval(x), \call]; \\
\call &\bowtie \hold[x, \eval(x)]; \\
\delta_i[\wait(x, \amb(y, \decide(z, v), v)), \wait(w, y)] &\bowtie \wait[\delta_i(x, w), z]; \\
\call &\bowtie \decide[\call, \epsilon]; \\
\epsilon &\bowtie \decide[x, x]; \\
@_i[x, \wait(y, \hold(@_i(x, y), \wait(v, w)))] &\bowtie \wait[v, w]; \\
\doublecap_i[\wait(x, y)] &\bowtie \wait[\doublecap_i(x), y]; \\
\sqcup_i[\wait(x, y)] &\bowtie \wait[\sqcup_i(x), y].
\end{align*}

\section{Remarks on efficiency}

Note that our solution is of proof-of-concept nature, and currently we do not directly aim at efficiency.
Still, our approach preserves some parallelism of Lamping's optimal algorithm, as there are potentially multiple $\eval$ agents operating simultaneously throughout the interaction net.
This is in contrast to the original token-passing net implementations~\cite{sinot1, sinot2}.

Our main concern was the capability of our system to handle corner cases for closed reduction~\cite{closed} and to reach the normal form of essentially $\lambda K$-terms.
During development and testing, some of the example terms we used were $(\lambda f x.K\, x\, f)\, \Omega$, $\omega\, (\omega\, (\omega\, (\lambda f x.f\, (f\, x))))$ where $\omega = \lambda x.x\, x$, and $(\lambda x.M)\, (Y\, I)$ where $x \not\in \fv(M)$ with various combinations of terms $M$ and fixed point combinators $Y$.

As our main test case, we used a complex $\lambda K$-term representing arithmetical expression ${3^3 - (2 + 2)!}$ with Church numerals and factorial defined via Turing's fixed point combinator.
The chosen $\lambda$-term was meant to cover most of the corner cases, and it also allowed us to collect comprehensive benchmarks.
For this test case, the total number of interactions was 2652687 out of which 2621262 are related to oracle nodes.
The number of interactions related to the waiting construct was 1182981 out of which 1159057 interactions were against oracle nodes.
In other words, the waiting construct occupied less than half of interactions, and the vast majority of its overhead is due to unoptimized oracle representation.

As shown in Section~\ref{waitsec}, the waiting construct propagates through the whole body of every abstraction applied to an argument.
If the waiting construct were modified to interact with oracle nodes in a more sophisticated way than just propagating through them, it could result in a run-time optimizer for oracle nodes and might possibly solve a long-standing issue about inefficiency of optimal reduction~\cite{lawall}.

\section{Read-back}
\label{readbacksec}

The process of decoding the normal form of an interaction net into the corresponding $\lambda$-term is called read-back.
Usually, read-back is described in prose, and only facilities external with respect to interaction nets are available.
Here we embed the read-back mechanism into interaction nets themselves.
Essentially, we dissolve application and abstraction agents into textual representation of the corresponding $\lambda$-term.

Through the rest of this section we will use the following notations: we denote the set of all $\lambda K$-terms as $\Lambda$, and $C[\phantom M]$ means a context, i.~e.~a $\lambda$-term with one hole, while $C[M]$ is the result of placing $M$ in the hole of the context $C[\phantom M]$.

In order to add the read-back mechanism to our interaction system, we further extend its signature:
$$
\Sigma = \Sigma_W
\cup
\{\top\}
\cup
\{a_M\ |\ M \in \Lambda\}
\cup
\{r_{C[\phantom M]}\ |\ \text{$C[\phantom M]$ is a context}\},
$$
with ${\ar(a_M) = 0}$ and ${\ar(r_{C[\phantom M]}) = \ar(\top) = 1}$,
the \textit{atom} agent $a_M$ encoding the textual representation of a $\lambda$-term $M$ and the \textit{read} agent $r_{C[\phantom M]}$ performing read-back in the context of $C[\phantom M]$.
In particular, agents $a_M$ make it possible to represent free variables in a given $\lambda$-term being encoded into interaction nets.
Note that our $\Sigma \setminus \Sigma_W$ extension is infinite as there has to be an agent for every $\lambda$-term and every context.

Let us recall that the original encoding in~\cite[pp.~41-42]{optimal} assumes no free variables in a $\lambda$-term being mapped into its initial encoding.
In order to allow free variables, we need to extend the mapping.
Specifically, while encoding $\lambda$-terms into our interaction system, we will distinguish their free variables from their bound variables.
So, let us mark all free variables in a $\lambda$-term $M$ using the following operation: ${M^\bullet \equiv M[\vec x := \vec x^\bullet]}$, where ${(\vec x) = \fv(M)}$.
Then, $\lambda$-term $M$ can be mapped to configuration
$$
\langle x\ |\ \eval(r_{[\phantom M]}(\top(x))) = y,\ [M^\bullet, y]\rangle,
$$
where the original encoding $[M, x]$ is extended with $[x^\bullet, y] = \{a_x = y\}$.

\begin{figure}[h]
\centering
\begin{minipage}{0.45\linewidth}
\caption{Read-back}
\label{readbackfig}
$$
\begin{tikzpicture}[baseline=(read)]
\matrix[row sep=0.5em]{
\inetcell(read){$r_{C[\phantom M]}$}[D] \\
\inetcell(lambda){$\lambda_i$}[U] \\ };
\inetwirefree(read.middle pax)
\inetwirefree(lambda.left pax)
\inetwirefree(lambda.right pax)
\inetwire(read.pal)(lambda.pal)
\end{tikzpicture}
\rightarrow
\begin{tikzpicture}[baseline=(read)]
\matrix[row sep=0.5em, column sep=0.5em]{
\inetcell(atom){$a_x$}[D] &
\inetcell(read){$r_{C[\lambda x.[\phantom M]]}$}[D] \\ };
\inetwirefree(atom.pal)
\inetwirefree(read.middle pax)
\inetwirefree(read.pal)
\end{tikzpicture}
$$
$$
\begin{tikzpicture}[baseline=(atom.above pal)]
\matrix[row sep=0.5em]{
\inetcell(appl){$@_i$}[D] \\
\inetcell(atom){$a_M$}[U] \\ };
\inetwirefree(appl.left pax)
\inetwirefree(appl.right pax)
\inetwire(appl.pal)(atom.pal)
\end{tikzpicture}
\rightarrow
\quad
\begin{tikzpicture}[baseline=(read)]
\inetcell(read){$r_{M\ [\phantom M]}$}[R]
\inetwirefree(read.middle pax)
\inetwirefree(read.pal)
\end{tikzpicture}
$$
$$
\begin{tikzpicture}[baseline=(read)]
\matrix[row sep=0.5em, column sep=0.5em]{
\inetcell(read){$r_{C[\phantom M]}$}[R] &
\inetcell(atom){$a_M$}[L] \\ };
\inetwirefree(read.middle pax)
\inetwire(read.pal)(atom.pal)
\end{tikzpicture}
\rightarrow
\quad
\begin{tikzpicture}[baseline=(atom)]
\inetcell(atom){$a_{C[M]}$}[L]
\inetwirefree(atom.pal)
\end{tikzpicture}
$$
\end{minipage}
\quad
\begin{minipage}{0.45\linewidth}
\caption{Conjecture}
\label{conjecturefig}
$$
\begin{tikzpicture}[baseline=(read.above pal)]
\matrix[row sep=1em, column sep=0.5em,ampersand replacement=\&]{
\inetcell(top){$\top$}[D] \\
\inetcell(read){$r_{[\phantom M]}$}[D] \\
\inetcell(eval){$\eval$}[D] \\
\inetcell[rectangle, inner sep=1em](atom){$[M^\bullet]$}[U] \\ };
\inetwirefree(top.middle pax)
\inetwire(top.pal)(read.middle pax)
\inetwire(read.pal)(eval.middle pax)
\inetwirecoords(eval.pal)(atom)
\end{tikzpicture}
\rightarrow^*
\,
\begin{tikzpicture}[baseline=(atom)]
\inetcell(atom){$a_N$}[U]
\inetwirefree(atom.pal)
\end{tikzpicture}
$$
\end{minipage}
\end{figure}

\newpage
Our read-back mechanism mainly consists of the three interaction rules that are shown in Figure~\ref{readbackfig}.
More formally in the interaction calculus, the rules related to read-back are as follows:
\begin{align*}
r_{C[\phantom M]}[x] &\bowtie \lambda[a_y, r_{C[\lambda y.[\phantom M]]}(x)], \quad \text{where $y$ is fresh}; \\
@_i[r_{M\ [\phantom M]}(x), x] &\bowtie a_M; \\
r_{C[\phantom M]}[a_{C[M]}] &\bowtie a_M; \\
r_{C[\phantom M]}[\doublecap_i(x)] &\bowtie \doublecap_i[r_{C[\phantom M]}(x)]; \\
r_{C[\phantom M]}[\sqcup_i(x)] &\bowtie \sqcup_i[r_{C[\phantom M]}(x)]; \\
r_{C[\phantom M]}[\wait(x, y)] &\bowtie \wait[r_{C[\phantom M]}(x), y]; \\
\eval[a_M] &\bowtie a_M; \\
\doublecap_i[a_M] &\bowtie a_M; \\
\sqcup_i[a_M] &\bowtie a_M; \\
\top[a_M] &\bowtie a_M; \\
\top[x] &\bowtie \doublecap_i[\top(x)]; \\
\top[x] &\bowtie \sqcup_i[\top(x)].
\end{align*}

Now, we believe that the following statement holds true.
It is still missing a formal proof.
However, no counterexamples have been found while experimenting with software implementation.

\begin{conjecture}
$\langle x\ |\ \eval(r_{[\phantom M]}(\top(x))) = y,\ [M^\bullet, y]\rangle \downarrow \langle a_N\ |\ \varnothing\rangle$
iff $N$ is the normal form of $M$.
\end{conjecture}
That is, the interaction net that encodes a $\lambda$-term $M$ in our interaction system reduces to normal form (if any) with no garbage and only one agent $a_N$ in its interface, $N$ representing the normal form of the encoded $\lambda$-term $M$; see Figure~\ref{conjecturefig}.

\section{Conclusion and further work}

To the best of our knowledge, optimal reduction had not yet been implemented using the approach of token-passing nets until this paper.
While our implementation of optimal reduction is rather a preliminary result, we still considered it worth sharing.
Also, we are not aware of any other existing implementations of read-back embedded into interaction nets and we covered it here in much more details than~\cite{mlc}.

In our opinion, further work should be directed to obtaining a proof for the conjecture given above if the latter holds true.
We believe that a proof may be achieved by verifying the following four aspects: non-interference with the original optimal reduction algorithm, the $\eval$ agents only triggering needed redexes, no possible deadlocks, and correctness of read-back.
Additionally, one needs to quantify how well our implementation preserves the parallelism of Lamping's optimal algorithm.
Finally, upgrading the waiting construct into optimization mechanism for oracle nodes and researching its efficiency looks promising for optimal implementations of functional programming languages.

\nocite{*}
\bibliographystyle{eptcs}
\bibliography{cite}
\end{document}